\documentclass[prd,aps,floats,tabularx,preprintnumbers,twocolumn]{revtex4}
\usepackage{graphicx, epsfig}

\begin{document}

\baselineskip 12 pt

\title{Do We Have Evidence for New Physics in the Sky?}

\author{Laura Mersini-Houghton$^\ast$}

\affiliation
{$^\ast$ Department of Physics and Astronomy, UNC-Chapel Hill, CB{\#}3255, Phillips Hall,Chapel Hill, NC 27599, USA}

\begin{abstract}
{\small}

Predicting signatures of string theory on cosmological observables is not sufficient. Often the observable effects string theory may impact upon the cosmological arena may equally be predicted by features of inflationary physics.
The question: {\it what observable signatures are unique to new physics}, is thus of crucial importance for claiming evidence for the theory. Here we discuss recent progress in addressing the above question.The evidence relies on identifying discrepancies between the source terms that give rise to large scale structure (LSS) and CMB, by cross-correlating the weak lensing potential that maps LSS with the CMB spectra.

\end{abstract}

\maketitle

\subsection{\bf Introduction}

Progress in string theory and precision cosmology has led to an ambitious investigation of the impact of new physics on cosmological observables.
The ultimate goal of any theory of quantum gravity is to predict a universe that looks like ours, at low energies. Currently string theory is the leading candidate for the anticipated new physics of the quantum gravity realm which, many suspect may hold the key to addressing the outstanding issues emerging in cosmology. Therefore the search for signatures of string theory in the cosmological arena is fully justified. The time to identify such signatures is right. Although searching for imprints of the new physics may be an ambitious goal, the tremendous progress of precision cosmology is bringing this goal within reach.

Testing the theory by observing signatures that are in agreement with the ones predicted by string theoretical models is neccessary but, it can not be considered sufficient evidence for the existence of the theory. The reason is because often the same signature can be predicted by alternative or standard model theories. For example, a predicted string signature on the temperature anisotropies spectrum can equally be predicted by inflationary physics, with a feature on the inflaton potential. We are then led to ask: {\it Can we uniquely identify and discriminate new physics signatures from features of inflationary physics in the presently available observables in the sky?} A positive answer would not only further motivate our search for new physics but it would also provide direct evidence for its existence.
Therefore, attempting to distinguish whether what is giving rise to these signatures are features of the inflaton \cite{easther,linde} or imprints of new physics \cite{efstathiou,katie,costa} is of crucial importance.This review discusses how the attempts, as recently proposed in \cite{steenlaura}, can be successful.
 We do not make use of the B-mode spectrum since measurement of the B-mode may not become available in a near future, although this spectrum, when available, can provide an independent handle for discriminating models through the consistency relation of the ratio of tensor to scalar amplitudes $r = A_t^{2} / A_S^{2}$ and cross-correlations with the $E, T$-modes.

\subsection{\bf Prelude: What kind of Universe do we live with?}

A very 'weird' one at present. A universe which, at present redshifts, $z \simeq 0-1$, and energy scales $H_0 \simeq 10^{-33}eV$, is: dominated by a mysterious component of energy, coined dark energy, driving it into an accelerated expansion \cite{de};and which, contrary to our theoretical expectations based on the inflationary paradigm, has large angle CMB perturbation suppressed \cite{wmap}.

Despite the cosmic variance limitations, let us take the WMAP CMB findings \cite{wmap} for large scales as true and consider the power suppression at low multipoles $l$ to be a real physical effect rather than a statistical fluke.

Dark energy domination and CMB power suppression of perturbations at  large angles, both occur, around the same redshift and energy scale $H_0 \simeq 10^{-33}$ eV. They both raise two disturbing questions: Why is their magnitude so small, (the tuning problem), and, why are they occurring now, (the coincidences problem). Besides, why do both coincidences occur at the {\it same} energy scale, our present Hubble radius $H_0$, \cite{ale}? It is tempting to speculate that too many coincidences in the present universe may indicate the emergence of a new low energy scale in physics of the order our Hubble scale $H_0 \simeq 10^{-33}$ eV. A consistent underlying theory should address both questions simultaneously. No such theory is yet known. A conservative approach would be to question whether we need new physics at all for addressing the current outstanding issues in cosmology. Distinct observational signatures would be the best way for resolving these doubts and for providing direct evidence of new physics if it exists.

\subsection{\bf Early Universe: CMB and LSS from the standard model of cosmology}

It is widely believed that at early times our universe underwent a period of inflation around the $GUT$ energy scale.
In the standard theory of inflation and physics of the early universe, large scale structure (LSS) and CMB perturbations are seeded by the primordial spectrum $\delta(k)$ that is traced back to quantum fluctuations of the inflaton field. Present day power spectrum is then obtained from the primordial spectrum by evolving it with the transfer function $T(z)$, $P(k)=T(z)\delta(k)$. Predicting the primordial spectrum and having it serve as the source that seeds all structure and perturbations in the universe is an amazing achievement of the inflationary paradigm. Let us briefly review the standard model expressions based on inflation, that relate $P(k)$ to the CMB spectra and LSS (mapped by the gravitational lensing potential), because they contain the information that will prove useful in discriminating inflaton features from new physics imprints.

The temperature perturbations given by the Sachs-Wolfe effect are

\begin{equation}
\label{temp1}
\Theta({\bf n}) = \frac{1}{3} \Phi(r_0,z_0) - 2\int_0^{r_0} \frac{d\Phi}{dr}(r,z(r))dr  \,\,
\end{equation}

where $\Phi(r,z)$ is the gravitational background potential and $r,z$ are the physical comoving distance and redshift respectively. The index $0$ denotes the value at last scattering surface. The gravitational potential $\Phi(r,z)$ is related to the growth factor of structure $G(z)$ through $\Phi(r,z) = (1+z)G(z)\Phi(r,0)$.Denoting the three dimensional matter power spectrum by $P(k)$ then the power spectrum for the CMB temperature anisotropies in a flat universe is

\begin{equation}
\label{temp2}
C_l^{TT} \simeq \int \frac{dk}{k^2} P(k) [\Theta_l (k)]^2 \,
\end{equation}

$C_{l}^{TT}$, (also $C_{l}^{LL, TL}$ defined below), are the $l-th$ Legendre cofficients in the expansion series of the angular power spectra $<\Theta({\bf n})^2>$, ($< L({\bf n})^2>, <\Theta({\bf n}) L({\bf n})>$ below). These spectra can be expressed as the convolution of the primordial spectrum with the appropriate transfer function and are obtained as solutions to the Boltzman equations, \cite{ma}.

An expression similar to Eqn.~\ref{temp2} gives the spectrum of the curl free E-mode of polarization of CMB photons, denoted by $C_l^{EE}$ \cite{ma}.

Notice that, through the power spectrum $P(k)$,( which is the primordial spectrum $\delta(k)$ evolved at present redshift $z$ with the transfer function), the same primordial $\delta(k)$ that produces the temperature anisotropies $C_l^{TT}$ also sources the polarization spectra namely the $E, B$ -modes and gravitational lensing of LSS. Gravitational lensing can be a powerful tool for mapping the background potential for LSS. The projected spectrum is described by the following potential 

\begin{equation}
\label{lens}
L({\bf n}) = -2\int_0^{r_0} dr \frac{r_0 -r}{r r_o}\Phi(r,z(r)) \,
\end{equation}

and the angular power spectrum of the lensing potential is

\begin{equation}
\label{lens2}
C_l^{LL} \simeq \int \frac{dk}{k^2} P(k) [L_l (k)]^2 \,
\end{equation}

where similarly $L_l (k)$ is the Fourier transform of $L({\bf n})$.
As shown by the expressions for the various spectra above, it is very useful to emphasize that  it is a predictions of the standard model of cosmology based on inflation, that {\it all spectra for CMB and LSS depend on the same source, the primordial spectrum}. The uniqueness of the source is the fact we will exploit in having a handle on identifying the non-inflationary contributions to the above spectra. The cross-correlation spectrum between $T,E,L$,Eqn.(~\ref{crossshear}), compares the sources of their respective autocorrelations to each-other.It can thereby identify any discrepancies with the expectations for having only the inflationary channel (the primordial spectrum) of the standard concordance cosmology. Cross-correlations are given by

\begin{equation}
\label{crossshear}
C_l^{TL} \simeq \int \frac{dk}{k^2} P(k)[ \Theta_l (k) L_l (k) ] \,
\end{equation}

A similar expression to Eqn.~(\ref{crossshear}) gives the cross-correlation spectrum $C_l^{TE}$ of T with E-mode.

The role of dark energy on the low multipoles $l$ is to enhance their power through the Integrated Sachs-Wolfe (ISW) effect. Instead the observed suppression of power is in disagreement with the ISW effect expected at large scales.Predictions of the concordance $\Lambda CDM$ cosmology, obtained from CMBFAST by the above equations, are shown by the solid line, in Fig.[1-7]. Might the disagreement at large angles of the $\Lambda CDM$ predictions with the observed $CMB$ spectra  indicate that clustering properties at large scales are different from the predictions of the standard theory and such that they compete with the ISW effect?

We show below how cross-correlations would reveal whether $LSS$ potential and $CMB$ spectra would deviate from their standard dependence on  the primordial spectrum due to new physics or conventional physics.If the source terms for the correlated spectra do not match as it is expected by the equations above then deviations may arise from contributions from noninflationary channels or a fundamental string scale imprinted on $C_l^{TL,TE}$ that may signal a breakdown of the conventional theory.

\subsection{\bf Signatures in autocorrelation spectra dont help!}
The vertical lines,in fig.[1-7], including the error bars marked by the height of the lines, represent real data released by WMAP and SDSS. Dark energy as a cosmic prior is also included, in accord with the combined data of LSS, SN, CMB.
As it can be seen in Fig[1-7], the concordance $\Lambda CDM$ model does not fit well with the CMB data at large angles. The integrated Sachs-Wolfe effect (ISW) of dark energy should have enhanced power at low multipoles $l$ since they correspond to perturbation wavelengths of the order our present horizon $H_0$ where $ISW$ takes full effect, while the WMAP findings, (as well as COBE experiment earlier) indicate power is suppressed at these scales.

Should we take this, together with the mystery of dark energy, as hints from data for new physics? Not yet. A power suppression at the observed wavelength scales can be equally obtained by introducing a feature, around the $60-th$ e-folding, on a $GUT$-scale inflaton potential. A bump on the inflaton potential which can violate the slow roll conditions can be as good of a candidate for giving rise to the observed CMB power suppresion\cite{easther,linde} as any of the more exotic possibilities associated with string theory and brane worlds, (although an $ad-hoc$ feature on the inflaton potential may not address the coincidence aspect of the suppression, namely, why the suppression occured only for such specific localized 'l' values, corresponding to perturbations modes with wavelength of the order our present Hubble radius $H_0 =10^{-33}eV$)). 
Let us investigate this possibility here: 

{\it Inflaton potential bumps around $k_{60}$ }. These models were studied in \cite{easther} by introducing a step in the inflaton potential $V(\phi)$
\begin{equation}
V(\phi) = \frac{1}{2}m^2 \phi^2 \left[1 + c\tanh(\frac{\phi - \phi_{step} }{d})\right] \ ,
\end{equation}

This potential has a step at $\phi = \phi_{step}$ with size and gradient controlled by the two parameters $c,d$ respectively. The step can be chosen to correspond to $k_{60}$. It was shown in\cite{easther} that besides departure from scale invariance and an oscillatory behaviour of the CMB spectrum this feature can produce significant effects at large scales. Thus using the observational input one can in principle constrain the parameters $\phi_{step},c,d$ such that this model comes in agreement with CMB $TT-$ spectrum at all scales, including the suppression of power at low 'l'.

The extreme case of the potential step function is to take a complete cuttoff of the primordial spectrum at $\phi_{step}$ thereby limiting the inflationary window to no more than 60 efoldings. This case is treated in \cite{linde} and the authors showed explicitly how this model can suppress quadropole power completely, and be in perfect agreement with the WMAP observations for $C_l^{TT}$. We refer to the class of single-field inflationary models with a feature on the inflaton potential or the primordial spectrum,like the examples above,as {\bf class A.}.

The lesson we learn from these two examples is that {\it inflationary models with a 'bump' around the $60$-th efolding can produce the same desired suppression of power at low multipoles, in fact an identical $TT$ or other autocorrelation spectrum for all $l's$,as compared to any of the more exotic models that would acquire the input of new physics}.Through these counter-examples we can conclude with the important point that : autocorrelations spectra (be it $TT,EE,LL$) do not help in finding evidence or testing our new physics theories. The reason is rooted in the fact that autocorrelations, Eqn.(~\ref{temp2}), compare the source to itself, (thus the word $auto$), which is why they can not identify where the modification of the source originate from nor be able to discriminate it against features in the primordial spectrum. {\it In short, independent of any observational findings at present or in the future, autocorrelations will not yield any new information on the origin of the features found and thus on the physics of the very early universe}.

\subsection{\bf Strategy: How to identify non-inflationary channels}

But there is hope in extracting new information about the high energy physics of the universe. Let us discuss in what way can we scrutinize signatures from the class of single-field inflationary models from imprints of new physics.

 The method of extracting this information relies crucially on comparing weak lensing potential with the sources that seed polarization and temperature spectra. It also makes use of the prediction of single-field inflation that the primordial spectrum is the only source that seeds all spectra of $CMB$ and $LSS$, as given by Eqn.(~\ref{crossshear},\ref{temp2},\ref{lens2}). {\it Therefore the modifications in spectra are overconstrained and do not allow any possibilities to simultaneously fit all data by e.g.simply introducing features on inflaton potentials}. A 'bump' on the inflaton potential will modify the primordial spectrum, thus $P(k)$. This means that if this feature is such that it suppresses $C_l^{TT}$ at large scales than it will automatically impose the same suppression, around the same multipole value 'l', for all other LSS and CMB spectra, $C_l^{EE}, C_l^{LL}$.

To illustrate this results, we investigated the class of single-field inflation with a feature, described above. Our analysis of the cross-correlations spectra $C_l^{TE,TL,EE}$, which have not been investigated for either model until now, are shown in Fig. [1,2,4]. It can be seen from the plots that all the auto and cross-correlation spectra for the class of inflationary models with a feature (dashed line graph) exhibit the same behaviour in the low 'l' regime since they are sourced by the same term. Due to the high correlation between LSS and CMB in class.{\bf A}, (since they share the same source), the expected correlation with LSS gravitational lensing potential is predicted to be around one for these models.
 Authors of \cite{Cline:2003ve} also noticed that an inflationary cuttoff could not accomodate the disrepancy between the $TT$ and $TE$ modes and tried to improve the fit by changing the optical depth. Results of our analysis in the next section indicate that exploiting the degeneracy among cosmic priors, e.g. changing the optical depth is not sufficient to bring this type of single-field inflationary models in agreement with data.( A similar analysis were later carried out for the multi-field inflationary models in \cite{waynehu}. Taking into account the astrophysical constraints on isocurvature channels,the conclusions obtained by the authors of \cite{waynehu} for the multi-field inflation class are in agreement with ours for the single field inflation class.{\bf A.}). Thus it is possible that some input of new physics may be required in order to bring agreement with data.

In \cite{steen} we proposed to use the combination of both correlation spectra $C_l^{TE}$ and $C_l^{TL}$  as our handle in identifying the origin of the imprints on observations while avoiding some of the degeneracies among cosmic parameters.Generically, string theory may offer more degrees of freedom than inflation by contributing to LSS and CMB with non-inflationary channels. For example, stringy moduli fields can couple to the matter sector and variations of these couplings can contribute as non-inflationary channels of perturbations.{\it The non-inflationary 'additional sources' provided by the underlying new physics, can only be revealed and identified by comparing the sources that seed CMB and LSS done by the cross-correlations}. We showed that autocorrelations cant reveal any information about the origin of the imprints. Here we explain how cross-correlations have the potential to reveal the information about contributions from non-inflationary channels of the new physics. In the cross-correlation spectra any features on the inflaton potential and their signature to LSS  and clustering properties at large scales would be very different from the effects of say an additional source to the primordial spectrum, provided by variations of moduli couplings carried over to the matter sector.Variations of moduli couplings could have a large impact onto the clustering properties of large scale structure and the polarization spectrum. 
If new physics of the early universe provides non-inflationary channels then their impact on deviations from the conventional large scale structure (LSS) and the clustering properties of matter at very large scale can be scrutinized against the overconstrained predictions for the inflationary spectra since in the former case clustering at large scales would be independent of the CMB thus we can predict that the correlation of CMB with LSS  must be very small. But in the latter the two are highly related to each other through their common dependence on the primordial spectrum, therefore $C^{TL}$ correlation is predicted to be of order one. 
This explains why discrepancies in the cross correlations reveal the complimentary information whether one source, (modified or not),is at works or additional sources are needed, thereby identifying the origin of the observed signatures.
 By applying our proposal to real data, predictions of single-field inflaton models with a feature are already in disagreement with the data, Fig.[1-4].Below we review a variety of possibilities of non-inflationary contributions to $LSS$ and $CMB$ spectra from string theory.

\subsection{\bf Non-inflationary channels from string theory}

There is no physical reason to expect the fundamental theory of the early universe to allow the existence of one degree of freedom only, the primordial spectrum, for structure and perturbations in the universe. Much of the recent work in string cosmology and phenomenology indicate that the hteory generically may provide more degrees of freedom than the standard cosmology. The hope is that some trace of them will survive in the low energy world due to the expansion of the universe\cite{katie}. A good place to start searching then is the cosmological arena. 
It is possible that the non-inflationary degrees of freedom may have contributed to LSS or CMB perturbations. Their contribution may become significant in the early or late time universe. For example couplings of various moduli fields predicted by string theory can vary in spacetime and thus contribute to perturbations.
Let us briefly review in this part some of the possible effects string theory may impact upon LSS or CMB sources, while trying to be as general and model-independent as possible. For consistency and in order to include the role of the ISW effect on perturbations,here we need consider only models that give rise to the observed present acceleration of the universe. The list below is representative and by no means exhaustive of all models and possibilities:

\subsubsection{ Possible String Imprints on a Low Energy World}

   {\bf i}) {\it Coupling of moduli to gravity/metric results in a modified Friedman equation}, at early or late times depending on the setup of the model, (e.g. \cite{cardassian,dgp,kogan,gregory,selftune,katie}).Let us generically denote the modification term by a power-law expression of the energy density $\rho$ scaled by the fundamental scale of the theory $M$ or by the brane tension $\sigma$ depending on the model, i.e a correction of the form $(\frac{\rho}{\sigma})^n$. If corrections to the Friedman equation grow with $\rho$, namely $n \ge 1$ then 

\begin{equation}
\label{early}
H^2 = G_N \left[ \rho \pm \rho (\rho /\sigma)^n \right] =\tilde G_N  \left [ \rho \right]   \ ,
\end{equation}

with 
\begin{equation}
\tilde G_N \simeq G_N \left[ 1 \pm (H_i^{1/2} / \sigma)^{n}\right ] \ ,
\end{equation}

$H_i$ denotes the Hubble parameter during inflation and $n \geq 1$ is some parameter of the theory depending on the specifics of the brane-world model. Since $\rho$ dilutes with time while the string scale $\sigma$ is a fixed parameter, then the linear term dominates over the higher order correction terms at late time thereby recovering the conventional Friedman equation. We refer to this class as {\bf case B.1.}.
If string modifications to general relativity or quantum field theory become non-negligible in the late-times universe, then corrections to the Friedman equation can parametrically be represented by the following equations
\begin{equation}
\label{late}
H^2 = \frac{8\pi G_N}{3} \left[ \rho + \frac{(1 - \frac{\rho}{\rho_c})H^{\alpha} }{H_0^{\alpha-2} }\right ] = \frac{8\pi \tilde G_N(z)}{3}\left [\rho \right] \ ,
\end{equation}

$\alpha$ is a parameter less than one and $H_0 \simeq 1/r_c$ is the present Hubble scale
given in terms of a crossover scale $r_c$ resulting from the higher dimensional nature of gravity in this string theoretical framework. (Examples of models in this class include \cite{cardassian},\cite{dgp},\cite{dgp,kogan,gregory},\cite{dvaliturner},\cite{selftune,stoicatye}). We refer to this class as {\bf case B.2.}.
Formally one can attribute the correction term, obtained after reducing to a $4-dim$ low energy world, to a rescaling modification of Newton's constant $G_N$
\begin{equation}
\label{eq:gn}
\tilde G_N= G_N \left [1 + (\frac{r}{r_c})^{(1-\alpha)} \right ] \,
\end{equation}

with $r$ comoving physical distance by expressing the energy density $\rho(z)$ as a function of $r(z)$.

 {\bf ii}) {\it Coupling of moduli to matter sector gives rise to short or long-range $5^{th}$ forces and nonlinear dispersion relations} in the matter sector.
Interaction of the moduli fields, $s$, with the matter sector, predicted generically by string theory, will give rise to $5^{th}$ forces, which are highly constrained by experiments. The carrier of the $5^{th}$ force can contribute to stringy modification of clustering properties.
Constraints from the $5^{th}$ force experiments may require many of these couplings to matter be forbidden, which can be achieved by appealing to global shift symmetry. However derivative couplings of moduli to $EM$ and $QCD$ fields, of the form

\begin{equation}
{\cal L}_{i} = g \frac{ s }{ M } F_{\mu\nu}{\tilde F^{\mu\nu}} \,
\end{equation}

 where $s$ is the moduli field, may not be suppressed by the shift symmetry. 

For the early times models, {\bf class B.1.}, fluctuations for the
light moduli fields are of the order the horizon size, therefore the $vev$ of the field becomes of order

\begin{equation}
<\delta s> \simeq H_i \,\,\,  \,  <s> \simeq 10^4 H_i =O(\sigma) \ ,
\end{equation}
 since  they should obey the perturbation constraint $\frac{<\delta s>}{ < s >} \leq 10^{-4}$.  
 
Derivative couplings of this type, say to 
 QCD strength tensor $G_{\mu\nu}$ or to the $EM$ fields, can give rise to short range forces. This interaction can also give rise to a dispersion relation for the background photons
\begin{equation}
\label{dispersion}
\omega^2 = k^2 + k \frac{g (<\dot s>)}{M} \ ,
\end{equation}

The induced dispersion relation for photons may be reflected in a shift in the polarization plane of photons which can affect the $C_l^{TE}$ correlation. Similar couplings to electrons may also induce fluctuations. Depending on the energy scales of the specific string model, if these couplings are relevant during recombination epoch, then they also provide a mechanism for inducing fluctuations in the electron number $n_e$, that enter the expression for Thomson scattering, as well as fluctuations in the optical depth $\tau$ due to the variation of the string coupling constant $g$ (in general a function of $<\dot s>, s, \sigma$ ).These fluctuations depend on $< \dot s >$ as follows

\begin{equation}
x =  \frac{\delta \tau}{\tau}    \simeq    \frac{\delta n_e}{n_e} 
\simeq \delta [\log(<\dot s>)] \,
\end{equation}
where we take $g \simeq O(1)$, and $x \simeq \frac{\dot H_i}{H M}$

In the {\it Late-time Universe} class of models, {\bf case B.2.}, the order of magnitude estimate for the range of values for $<\delta s>, <s>$ are very different as seen in Eqn.(8,10). The $s$-field  for case {\bf B.2} has a range
of force of order the crossover scale $r_c$ taken to be roughly our present Hubble radius $r_c=H_o^{-1}$. Therefore coupling to these moduli gives rise to {\it long-range} $5^{th}$ forces. 
 Due to the severe constraints on the long-range 5th force experiments and variation
of the $\alpha$ constant, coupling to the QCD fields are forbidden in this group or have to be very
highly suppressed, \cite{5force}.

Similarly, derivative couplings to the $EM$ field are not suppressed by the shift symmetry

\begin{equation}
{\cal L}_{i} = g \left ( \frac{ \nabla_{\mu} s }{M} ) \right) j^{\mu} \ ,
\label{emcouple}
\end{equation}

and can give rise to a dispersion relation for photons, obtain by the interaction Eqn.(~\ref{emcouple})

\begin{equation}
\omega^2 =k^2 + k \frac{g}{M}  (<\dot s>) \ ,
\end{equation}

Here $<\dot s> $ may be a positive quantity since the Hubble parameter $\dot H_0 \geq 0$ at low redshifts due to the accelerated expansion of the universe. 

The nonlinear behavior of the dispersion relation becomes important when the second term is dominant, $k < \frac{g <\dot s>}{M} \simeq g /[1+z]$ since roughly 
\begin{equation}
\label{sflux}
<\dot s> \simeq  \frac{H M}{[1+z]} \,\, < s > \simeq M \log[1+z]
\end{equation} 

The shift in the angle of polarization plane for photons is e.g. $\delta \theta \simeq (g/[1+z])$ where z is the
redshift (see also \cite{balaji}). The optical depth variations for these scenarios are very small, with an order of magnitude 

\begin{equation}
\frac{\delta \tau}{\tau} \simeq \delta (\log [H_{equality}]) \ ,
\end{equation}

where
equality means the mater radiation equality time, at low redshifts. This estimate follows from the relation $\delta [\log(<\dot s>] \simeq \delta \log H(z_e)$. The latter is a consequence  of energy conservation equation.


{\bf iii)} {\it Modulated string perturbations}. String coupling constants, $g$ are functions of moduli fields. Varying moduli vev's give rise to varying coupling constants which in turn can generate a new channel of noninflationary perturbations, example \cite{dvalikofman}

\begin{equation}
\label{varg}
\frac{\delta g(s)}{g(s)} \simeq \frac{\delta T}{T} \simeq \frac{\delta \Phi[r,z]}{\Phi[r,z]}
\end{equation}

 This effect of modulated perturbations was treated in \cite{dvalikofman}where a suppressed $B$-mode was offered as a signature on the spectra from the varying of coupling constants but the B-mode data is not available at present.

{\bf iv)} {\it Different clustering at large scales}. Any of the string imprints in $(i-iii)$ above can change the clustering properties of large scales by modifying the  background gravitational potential $\Phi$ for perturbations, e.g Eqn.~(\ref{varg}). Moduli coupling to matter sector and the varying coupling constant may contribute with a new channel of modifications to the gravitational background potential, of the form

\begin{equation}
\Phi \simeq \Phi_{inflation} + \phi \left [g, \frac{<\delta s>}{<s>}\right] \ ,
\end{equation}
 The correction to the Newtonian potential of the background, $\phi [ g, \frac{<\delta s>}{<s>} ]$, can change the clustering properties at very large scales. The details of $\phi [g, \frac{<\delta s>}{s}]$ depend on the specific string model considered.They can be generated by the mechanism described in \cite{dvalikofman}. They can also be generated by the mechansim described in \cite{steen} where the carrier of the long range $5^{th}$ force, resulting from coupling of moduli to the matter sector, introduces a correction $\phi [g, \frac{<\delta s>}{s}]$ to the gravitational potential. For a coupling constant $g$ proportional to some power $(1- \alpha)$ of moduli, Eqn. (\ref{sflux}), in case {\bf B.2}, becomes

\begin{equation}
\label{phi}
\frac{\delta g}{g} \simeq \frac{<\delta s>}{<s>} \simeq \frac{\delta \Phi}{\Phi}=\frac{\phi}{\Phi}=\left(\frac{r[z]}{r_c}\right)^{1-\alpha}
\end{equation}

where  $r_c$ denotes the combination $r_c = g/M $. These type of modifications to $\phi(r,z)$ and their predictions for clustering clustering at LSS and cross-correlation spectra, were analysed and the plots given in Fig[3-7] are described in the next section.

The correction term can be identified and deduced from $C_l^{TL}$, Eqn.~(\ref{temp1} - \ref{lens2}). If the expression for $\phi[g, \frac{<\delta s>}{<s>}]$ contains a new string scale then we could deduce the emergence of this 'new string scale' related to the coupling constant $g$.  A coupling constant that varies slowly over cosmological scales, (e.g. oscillating every one Hubble time), is such an example of a significant contribution to clustering at large distances. See Fig.[4,5].

This analysis can also be used to discriminate among models within Class {\bf B}, since while class {\bf B.1} of Early Times modifications distorts the LSS and CMB spectrum through modifications of the primordial spectrum, the class {\bf B.2.} of Late Times modification will manifest its signature as a modification to the ISW effect as well.


 Unlike the single field inflation class.{\bf A.}, as described above, it is natural for string theoretical models of cosmology, class {\bf B.1., B.2}, to provide more degrees of freedom and more than one source for seeding LSS and $T,L,E$ spectra.

Let us scrutinize these cases one by one below. Our analysis is carried out with the data already available and the one expected in a near future from weak lensing measurements for large scales.

\subsection{Data Analysis}
A rigorous derivation of realistic cosmological scenarios is yet to be obtained from string theory. Lacking a consistent effective theory derived from the fundamental one, means that we can not predict their perturbation equations or the modifications to all Einstein equations.In principle perturbations of string motivated models may be quite different from the standard linear perturbation theory, although the literature in this field assumes that standard perturbation theory is a good approximation. The analysis of real data described below assumes the validity of linear perturbation theory.


\subsubsection{Large Scale Structure (LSS) and Cosmic MicroWave Background (CMB)}

At present there are two large galaxy surveys of comparable size, the
Sloan Digital Sky Survey (SDSS) \cite{Tegmark:2003uf,Tegmark:2003ud}
and the 2dFGRS (2~degree Field Galaxy Redshift Survey) \cite{2dFGRS}.
This analysis uses data from SDSS and only data points on scales larger than $k = 0.15 h$/Mpc in order to avoid problem with non-linearity.

The CMB WMAP experiment has reported data only on $C_l^{TT}$ and $C_l^{TE}$
as described in
Refs.~\cite{Spergel:2003cb,%
Verde:2003ey,Peiris:2003ff}.  We have performed our
likelihood analysis using the prescription given by the WMAP
collaboration~\cite{wmap,Spergel:2003cb,%
Verde:2003ey,Peiris:2003ff} which includes the
correlation between different $C_l$'s. Foreground contamination has
already been subtracted from their published data.

\subsubsection{Likelihood analysis}

The set of cosmological parameters, other than those related to modified gravity, is the minimum standard model with 6
parameters as given in Tab.1.for  geometrically flat models $\Omega = 1$. The normalization of both CMB and LSS spectra are taken to be free and unrelated parameters.Theoretical CMB and matter power spectra are calculated by using CMBFAST.

\begin{table}
\begin{center}
\begin{tabular}{lc}
\hline
parameter & prior\cr
\hline
$\Omega_m$ & $0-1$ (Top hat) \cr
$h$ & $0.5-1.0$ (Top hat) \cr
$\Omega_b h^2$ & $0.014-0.040$ (Top hat) \cr
$\tau$ & $0-1$ (Top hat) \cr
$b$ & free \cr
\hline
\end{tabular}
\end{center}
\caption{The different priors on parameters
used in the likelihood analysis. Parameters related to modified gravity are not tabulated here.}
\label{table:prior}
\end{table}

The suppression of TT power at large scales for Class {\bf A, B.1} of models with modified primordial spectrum has been treated in literature. We analyzed their polarization spectrum and $TE$ correlation. Our analysis shows that the $T, E$ low multipoles cannot simultaneously be fitted by a break in the primordial power spectrum. The main reason is that modifications to the $TT$ and $TE$ power spectra are overconstrained by the primordial spectrum to change in the same direction as the initial spectrum. Rhese results are shown in Fig.[1-3], ({\it dashed-line}). Theoretical predictions of single field inflation with a feature thus show disagreement with the findings of WMAP data of a low $TT$ power but a {\it high $TE$ power} at small multipoles. Exploiting degeneracies among cosmis priors,e.g. adding a high optical depth, can not fit their spectra to observation.

The likelihood analysis of late-time models, {\bf Class.B.2} was done by taking Deffayet et al. \cite{Deffayet:2001pu,Deffayet:2002sp} as an illustrative example. There is no change in the perturbation source terms for this model. (Other examples for Class.{\bf B.2.} can be found in \cite{cardassian} and \cite{dvaliturner})where the Friedman equation is modified directly). Although it provides a good fit to CMB and LSS data, it is a poor fit to the combination of all available data including measurements of the Hubble parameter. Parameter fitting to more general modifications of the Friedmann equation can for instance be found in \cite{elgaroy}

\begin{figure}
\includegraphics[width=80mm]{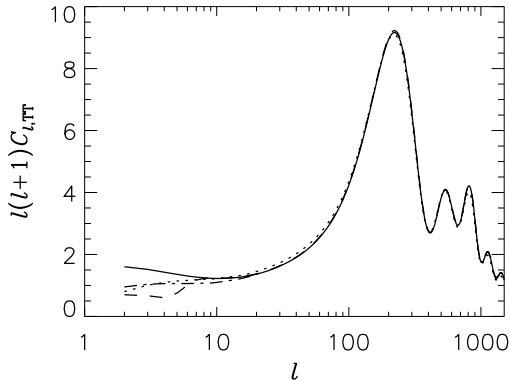}
\caption{$TT$ spectra of the various models. The full line is the standard $\Lambda$CDM model, the dashed line has a modified primordial power spectrum, the dot-dashed a modified perturbation source term (case {\bf b}). Finally, the dotted line has a modified Friedman equation and modified source terms (case {\bf a}). Normalization is arbitrary.}
\label{fig:cl}
\end{figure}

As discussed in \cite{steenlaura} a change in the effective strength of gravity can also leads to a modified background Newtonian potential and  modified clustering properties at large scales. Both, the modified Friedman equation and, the modified $G_N$ are effective equations obtained from the underlying theory. Then there is no fundamental reason why modifications for CMB spectra and LSS clustering can not occur independently. A likelihood analysis performed for this new approach which, besides the modifications to the Friedman equation , Class {\bf B.2} \cite{cardassian, dvaliturner, Deffayet:2001pu},contains also a  perturbation source modified according to $G_N$ in Eq.~(\ref{eq:gn}) with $\alpha=0$ corresponds to {\bf case a} in the plots, Fig[1-7]. For comparison to {\bf case a}, the likelihood analysis for the case when only the perturbation source terms are modified,(but with no modification to the Friedman equation),is shown and it corresponds to {\bf case b} in the plots. In both cases $r_c$ is a free parameter.  
The best fit models for {\bf cases a} and {\bf b} are shown in Fig.~\ref{fig:cl}. Both have $k_c = 2\pi/r_c \simeq  3 \times 10^{-6} \,\, h$/Mpc.
The standard $\Lambda$CDM model, and inflationary models with 
broken scale invariance are shown for comparison. The latter is a concordance $\Lambda$CDM model with no power at $k < 5 \times 10^{-4} \,\, h$/Mpc, the same as the model studied in Ref.~\cite{kesden}.On small scales all the models are clearly indistinguishable, but at low multipoles there are significant differences.

\begin{figure}
\includegraphics[width=80mm]{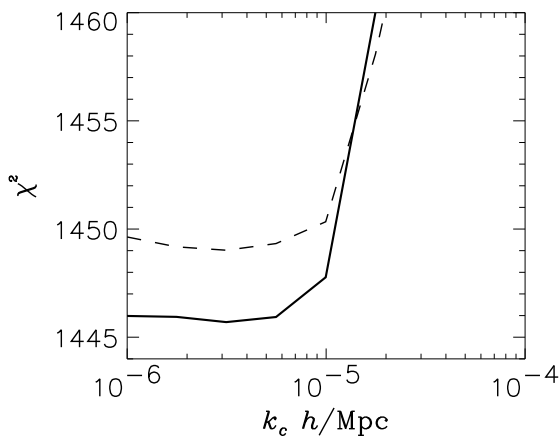}
\caption{$\chi^2$ as a function of $k_c$. The full line is for case {\bf a} and the dashed for case {\bf b}.}
\label{fig:chi2}
\end{figure}

\begin{figure}
\includegraphics[width=80mm]{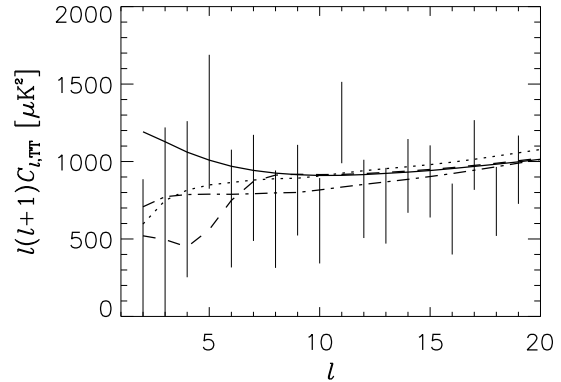}\\
\includegraphics[width=80mm]{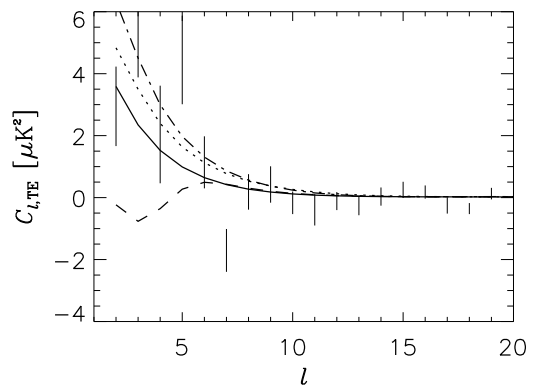}
\caption{$TT$ and $TE$ spectra of the various models. The full line is the standard $\Lambda$CDM model, the dashed line has a modified primordial power spectrum, the dot-dashed modified perturbation source terms (case {\bf a}). Finally, the dotted line has a modified Friedman equation and modified source terms (case {\bf b}). The data points are those measured by WMAP}.
\label{fig:cl2}
\end{figure}

$TT$ and $TE$ large scale spectra are shown in Fig.~\ref{fig:cl2}. As can be seen, models with modified gravity are able to simultaneously produce less $TT$ power and more $TE$ power, and are in slightly better agreement with data than the standard model. The model with broken scale invariance is clearly a quite poor fit.
Fig.~\ref{fig:chi2} which shows $\chi^2$ as a function of $k_c$ for the two cases also slightly favors {\bf case a} models.

It is worth noting that in Class {\bf B.2}, if $r_c$ is fixed by the acceleration of the universe,namely the present Hubble radius to $k_c = 2\pi/r_c \simeq 10^{-3}$ the $\chi^2$ for the best fit model is roughly $\chi^2=1.6 \times 10^4$, i.e.\ it is completely ruled out.

\begin{table}
\begin{center}
\begin{tabular}{lccc}
\hline Model & $\chi^2$ & $\Omega_m$ & h \cr \hline 

$\Lambda$CDM & 1449.5 & 0.30 & 0.69 \\

Case B & 1449.0 & 0.31 & 0.69 \\

Case A & 1445.7 & 0.42 & 0.58 \\
\hline

\end{tabular}
\end{center}
\caption{Best fit $\chi^2$ for the standard 6 parameter $\Lambda$CDM model, as well as modified gravity cases {\bf a} and {\bf b}. We also show the best fit values of $\Omega_m$ and $h$.} \label{table:parameters}
\end{table}

\subsubsection{Clustering at large scales: Direct modification of the gravitational potential}
As discussed above, the data analysis shows that a new channel of contributions, besides the inflationary one, for the clustering properties at large scales, of order the present horizon, seems the most promising class of models.
\begin{figure}
\includegraphics[width=80mm]{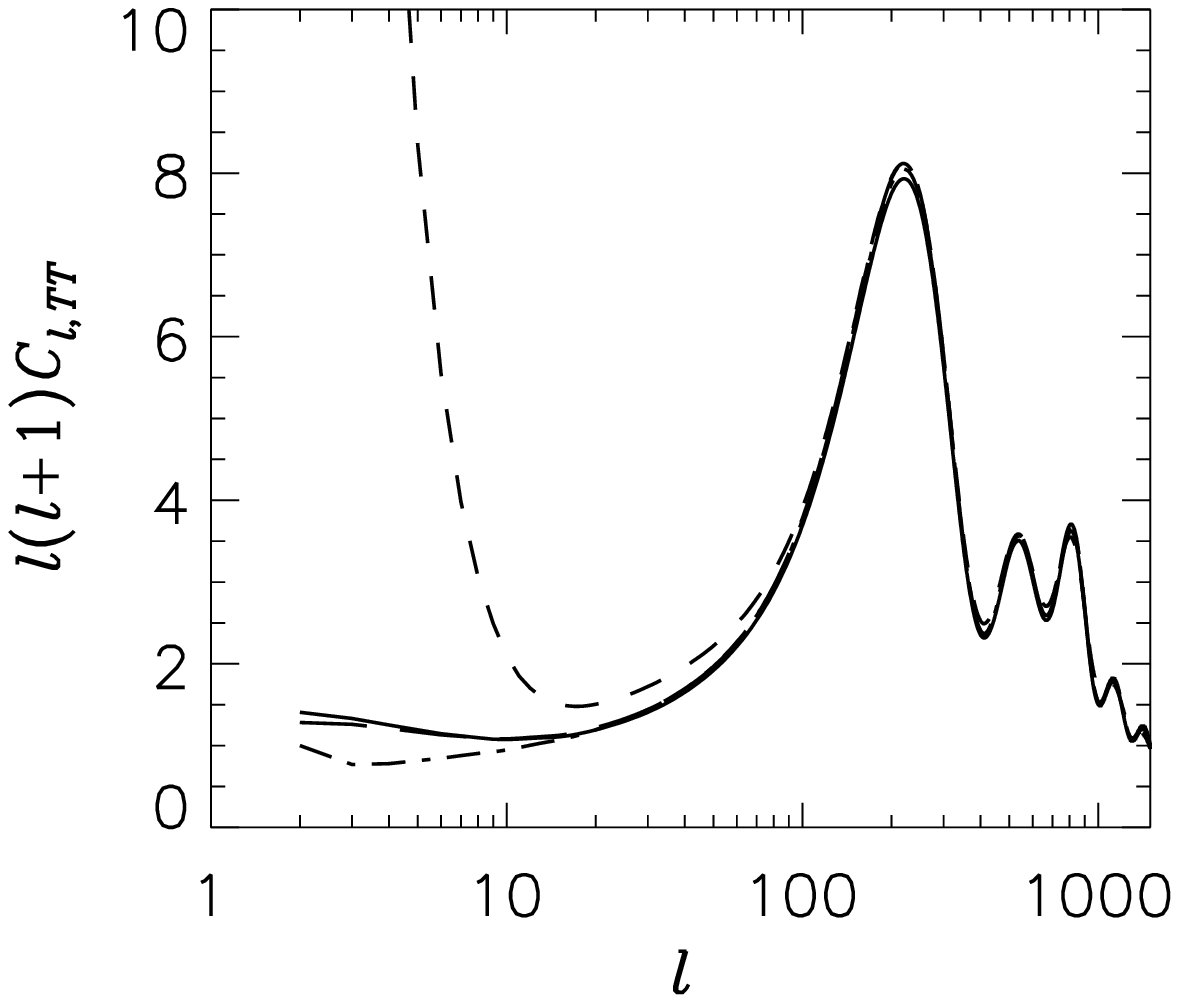}\\
\includegraphics[width=80mm]{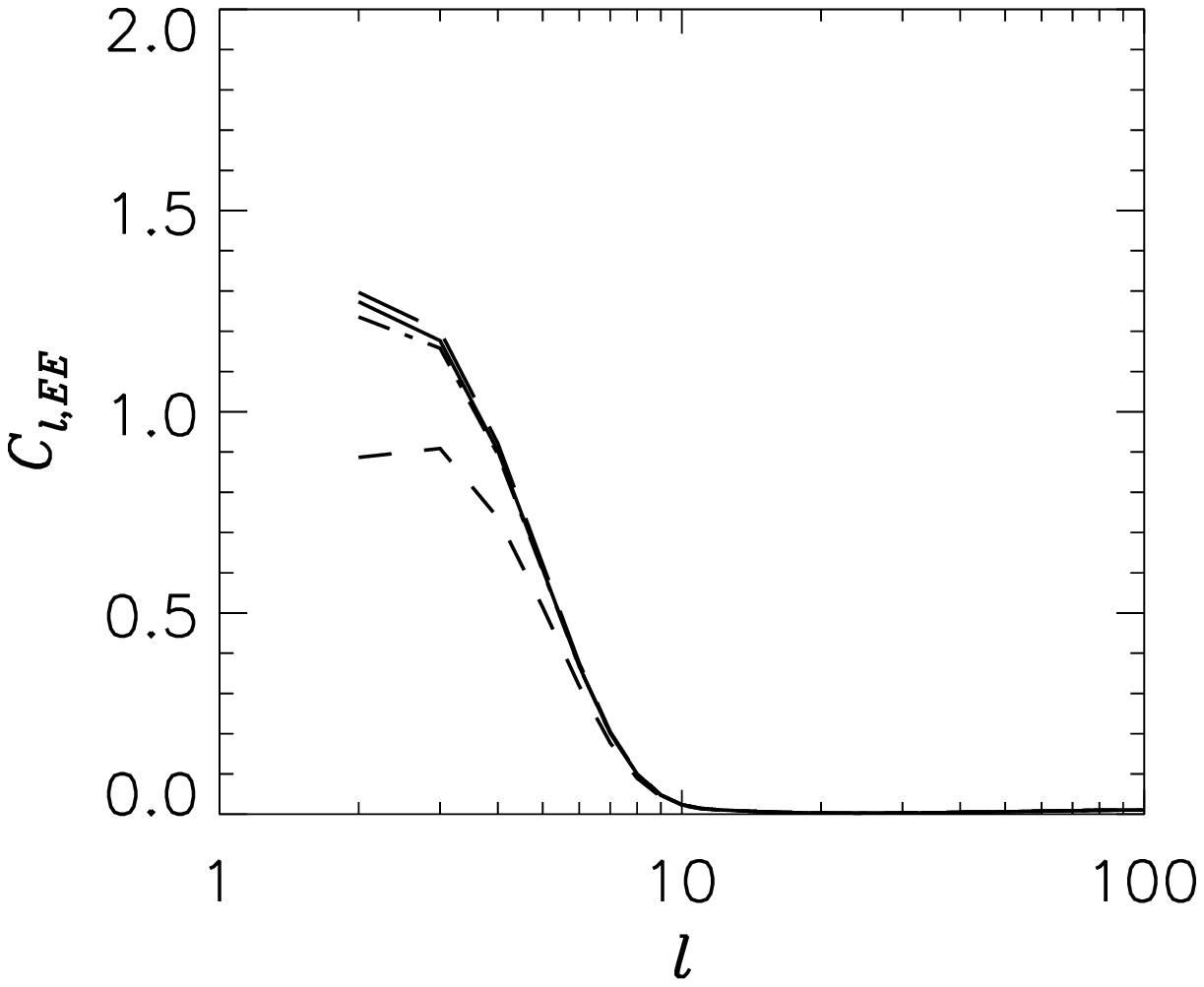}
\caption{$TT$ and $EE$ spectra of models with a modified $\phi$ according to Eq.~(\ref{phi}). The curves are for $\alpha=0$ and various values of $k_c = 2\pi/r_c$. The full line is for $k_c=0$, the long-dashed for 
$k = 10^{-6} \,\, h$/Mpc, the dot-dashed for 
$k = 10^{-5} \,\, h$/Mpc, and the dashed for
$k = 10^{-4} \,\, h$/Mpc. Normalization is arbitrary.}
\label{fig:mod1}
\end{figure}

\begin{figure}
\includegraphics[width=80mm]{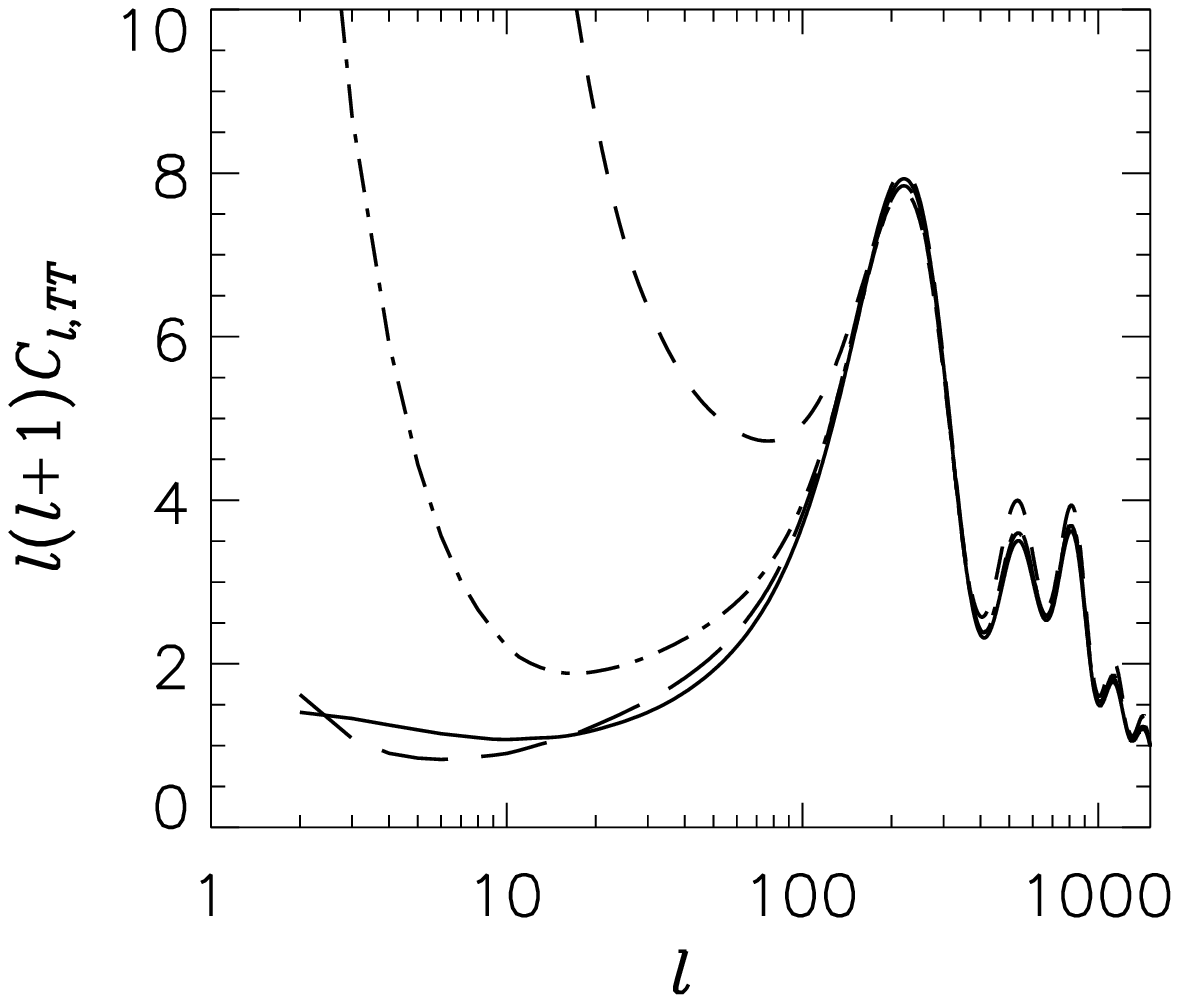}\\
\includegraphics[width=80mm]{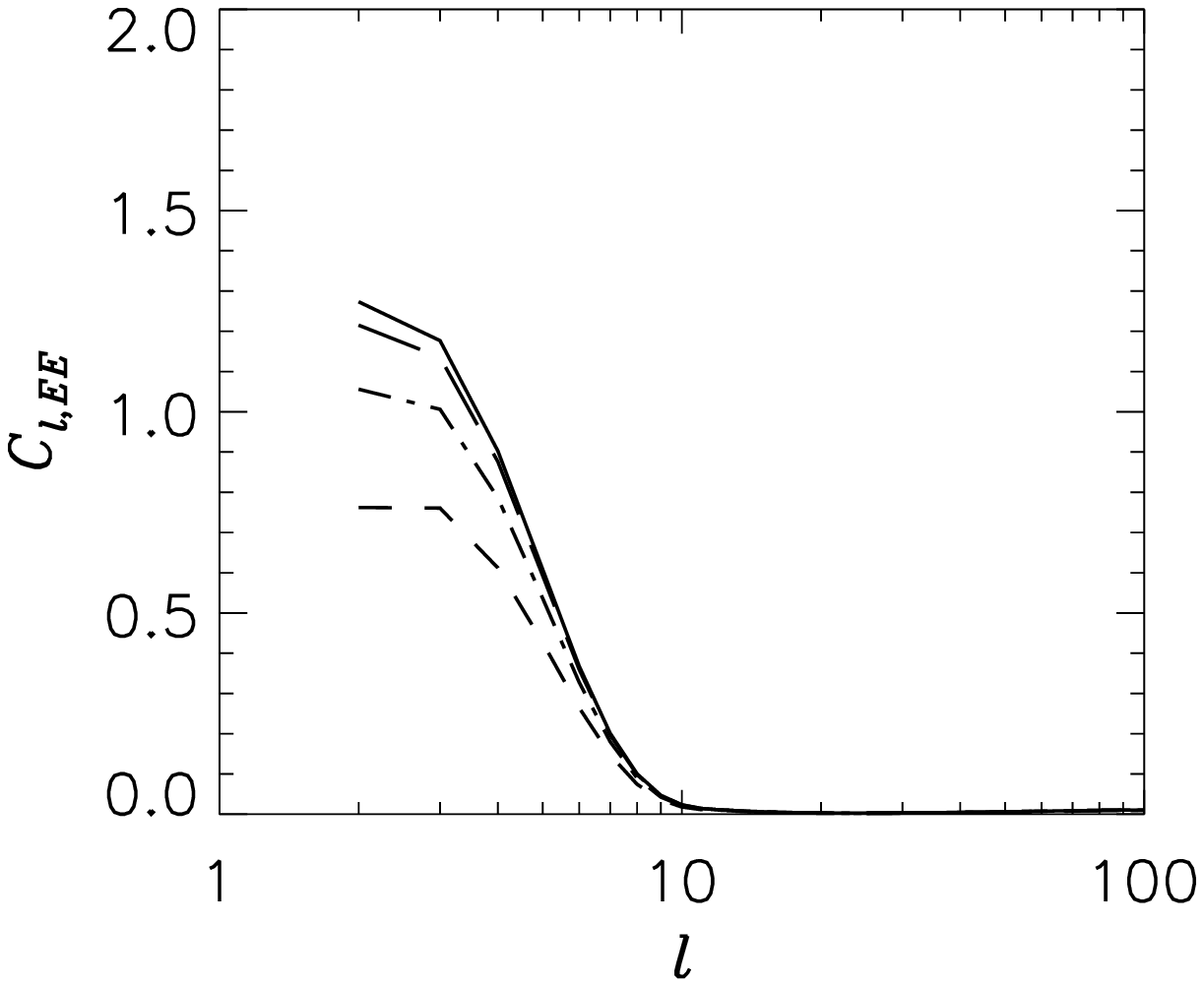}
\caption{$TT$ and $EE$ spectra of models with a modified $\phi$ according to Eq.~(\ref{phi}). The curves are for $\alpha=0.5$ and various values of $k_c = 2\pi/r_c$. The full line is for $k_c=0$, the long-dashed for 
$k = 10^{-6} \,\, h$/Mpc, the dot-dashed for 
$k = 10^{-5} \,\, h$/Mpc, and the dashed for
$k = 10^{-4} \,\, h$/Mpc. Normalization is arbitrary.}
\label{fig:mod2}
\end{figure}

\begin{figure}
\includegraphics[width=80mm]{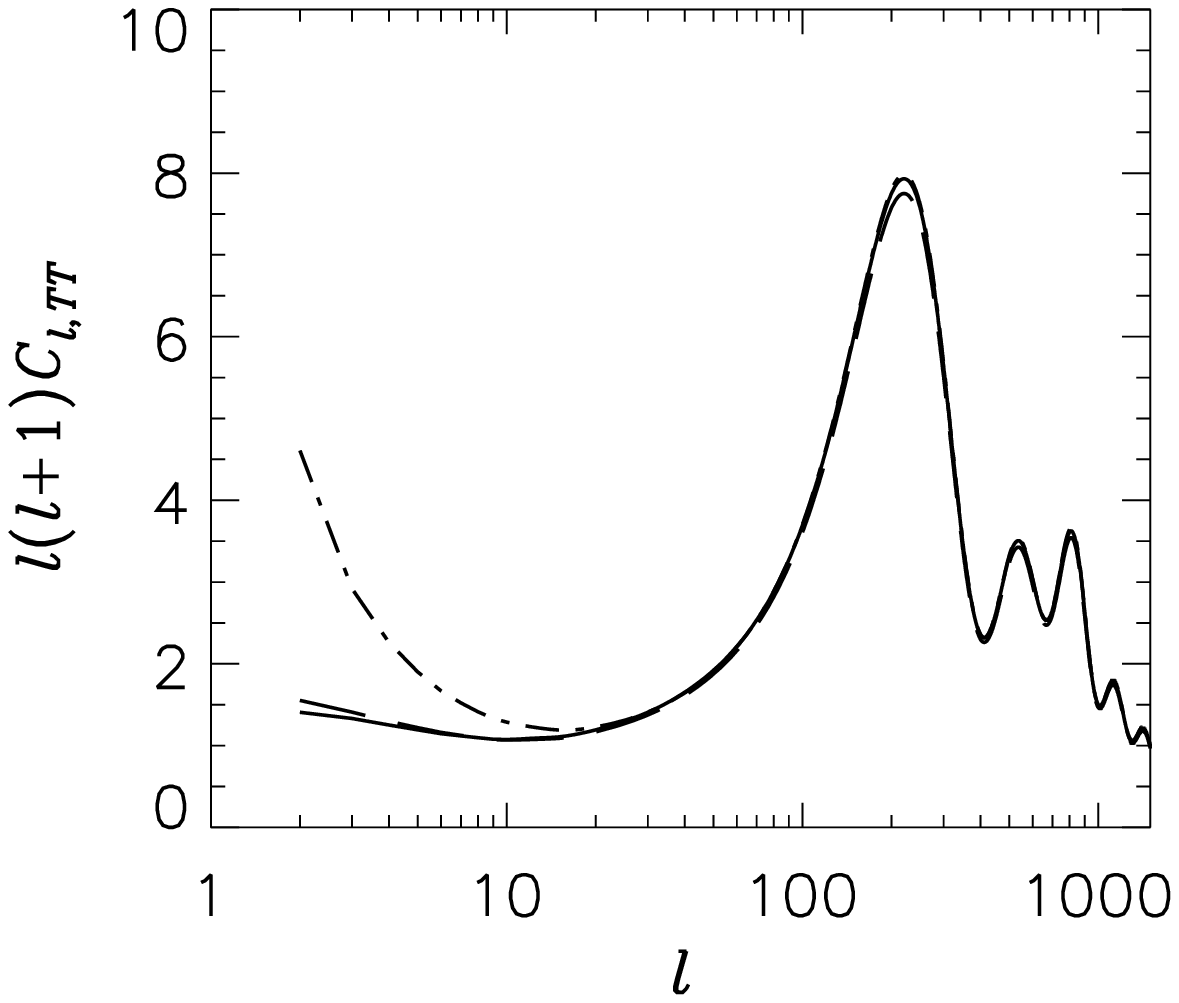}\\
\includegraphics[width=80mm]{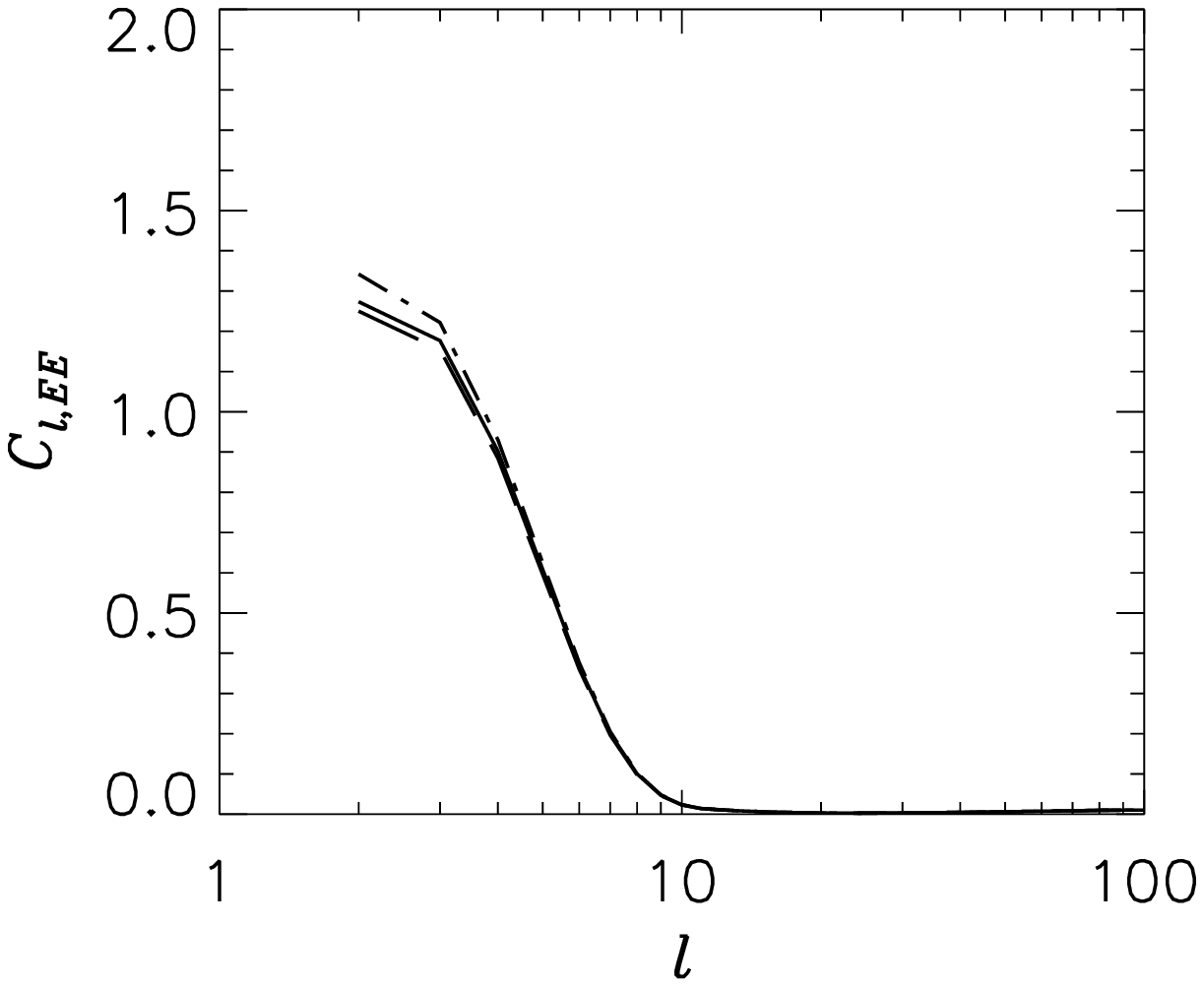}
\caption{$TT$ and $EE$ spectra of models with a modified $\phi$ according to Eq.~(\ref{phi}), but with NEGATIVE sign. The curves are for $\alpha=0$ and various values of $k_c = 2\pi/r_c$. The full line is for $k_c=0$, the long-dashed for 
$k = 10^{-6} \,\, h$/Mpc, and the dot-dashed for 
$k = 10^{-5} \,\, h$/Mpc. Normalization is arbitrary.}
\label{fig:mod3}
\end{figure}

Fig.[4,5] shows the analysis for the case of a direct modification to clustering properties at large scales by stringy corrections to the background gravitational potential. The specific case shown in the plot is the correction $\phi$ of Eqn.~(\ref{phi}) which, although of different origin, resembles the correction to Newton's constant of the DGP model and thus modifies the LSS source similarly. Interestingly an agreement with data indicates that the turnover scale, which is proportional to the string coupling constant $g(s)/M \simeq r_c$ has to be larger than the present Hubble radius. (A similar result applies to the DGP model, their turnover scale, $r_c^{-1}$, has to be larger than $H_0$ to reach agreement with data. Authors of \cite{dgp} take the scale $r_c \simeq  H_0$ in order to explain the acceleration of the universe but this choice is disfavored by the $LSS+CMB$ data). From Fig.~\ref{fig:mod1} and \ref{fig:mod2} it can be seen that the suppression of large scale temperature power spectrum for this class of models belonging to {\bf case a} is in agreement with the WMAP findings. Contrary to predictions of Class.{\bf A} of inflationary models with a feature which predict a suppressed polarization and $TT$ spectrum simultaneously, the $E$-polarization power spectra of models with modified clustering properties at large scales, is not suppressed or modified in any way. The reason is that the polarization anisotropy is solely related to physics around the epoch of recombination and to possible reionization. For illustration, Fig.~\ref{fig:mod3} shows the spectra of a direct modification of the background potential $\phi$ for the case when the correction term is negative in Eqn.~(\ref{phi}). This can arise for example from a deconfining $5^{th}$ force. The plot indicates that these models which reduce the background potential at large scales through negative contributions from the correction $\phi$ are poor fits to the observational data.

\subsubsection{Future data}

\begin{figure}
\includegraphics[width=80mm]{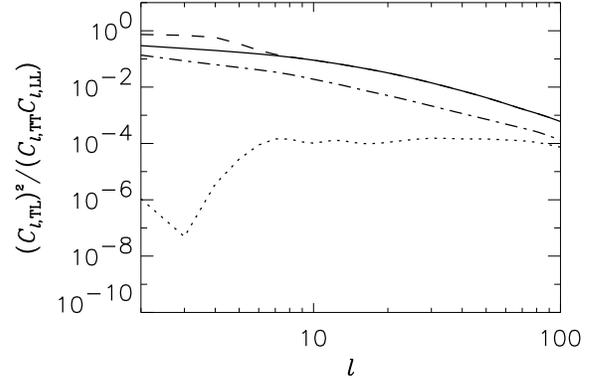}
\caption{Cross correlation between shear and temperature. The full line is the standard $\Lambda$CDM model, the dashed line has a modified primordial power spectrum, the dot-dashed modified perturbation source terms (case {\bf b}). Finally, the dotted line has a modified Friedman equation and modified source terms (case {\bf a}).}
\label{fig:cross}
\end{figure}

Future CMB observations can at most improve the very large scale precision moderately because of cosmic variance limitations.
However, future data may be able to distinguish among the various types of models as discussed below. We proposed to address this issue in \cite{steenlaura} by using large scale cross correlation between shear and temperature fluctuations, $C_l^{TL}$. Fig.~\ref{fig:cross} once data from $LSS$ surveys is available. The class of inflationary models with modified primordial spectrum produces increased $TL$ correlation at large scales, Fig.[7]. The opposite prediction is true for {\bf case a} with non-inflationary modifications of the clustering properties at large scales. The latter has no correlation between temperature and cosmic shear due to the existence of a new non-inflationary channel, besides the modified primordial spectrum. Therefore predictions for the $TL$ correlation are nearly zero when new non-inflationary channels contribute either to clustering or perturbations, case {\bf a},as compared to the case with only one (inflationary) source, modified or not. In the latter the $TL$ correlation is order one. Case {\bf b} (only modified source terms) resembles the standard model in terms of the $TL$ spectrum since this class does not add an independent new channel to either clustering or perturbations.This difference in the temperature shear correlation should be clearly visible in future weak lensing surveys which would find no detectable $TL$ power for case {\bf a} models and order one power for the $one-$source only models. We expect that {\it future weak lensing experiments at very large scales will play a major role in providing new information for cosmology and direct eveidence for new physics, if it is there}.

Upcoming WMAP $E$-polarization measurements will also provide an independent test to models with stringy modifications of gravity or clustering  at late times, since these models predict little or no change in $E$-polarization compared with standard $\Lambda$CDM. This is a testable prediction because models with early time modifications to the power spectrum directly change the primordial spectrum and therefore influence $C_l^{EE}$,while models with late-time string effects do not influence the polarization spectra.

A noteworthy point, not treated here, is the study of observational constraints on modifications of the photon dispersion relation, Eq.~(\ref{dispersion}) that can lead to changes in both the power spectra and in the observed Planck spectrum of the CMB, similar to a chemical potential (see \cite{dispersion} for bounds on this possibility). This point merits further study.

\section{Discussion}

Recent advances in precision cosmology and especially the discovery of the acceleration of the universe, has presented cosmology and modern physics with serious challenges. A model that addresses these challenges at a satisfactory level does not exist yet. By now, many believe that an explanation of the observed anomalies may be rooted in the unknown new physics. Skeptics could equally argue that conventional physics has been as effective as the new physics models in accomodating the data. Besides, the latter seems better motivated. Predicting observable imprints of string theory, while neccessary it is not sufficient.  Therefore the question: {\it how can we distinguish new physics imprints from features of the concordance model} is of crucial importance. This question is equivalent to identifying observable effects that are unique to new physics. String theory is the leading candidate for quantum gravity, thus a positive answer to the above offers a direct evidence for string theory.

Here we reviewed the proposal of \cite{steenlaura} that addresses this question based on the analysis of cross-correlations at large scales, of $T$ with polarization $E$ and lensing potential $L$. The method described here can potentially provide the first direct evidence for the existence of new physics, as summarized below.

The autocorrelation spectra, $TT,EE,LL$ can neither discriminate between the sources that seed LSS and CMB spectra nor be able to identify a non-inflationary channel because they compare the source to itself. Independent of the potential signatures predicted by string motivated models or observed in autocorrelations, autocorrelations are incapable of providing evidence for new physics, even if it is there.

The $C_l^{TE}, C_l^{TL}$ correlations can uniquely identify imprints of stringy modifications and discriminate these signatures from features on the inflaton potential because they cross-examine the mismatch in the sources that seed the spectra. A comparison of $C_l^{TT}, C_l^{TE}, C_l^{TL}$ reveals whether the sources that seed these spectra differ from one another. Analysis of present data indicates that at least two different sources or equivalently two different scales are required in order to reach agreement with observation. One of these scales is related to the significance of the ISW effect on CMB produced by present acceleration and, the other one is related to modifications of horizon size clustering properties,i.e. at scales much larger than the recombination era in order to preserve the observed high $C_l^{TE}$. 

Models with features on inflationary potentials (Sect.2 case {\bf A}) or of Early-Time brane-worlds (Sec.2 case {\bf B.1}) with modified Friedman equation which change {\it only} the primordial spectrum, but do not provide additional sources for modifying the late-time evolution and large scale clustering
in general are not able to explain the {\it high} $C_{TE}$ and the {\it low} $C_{TT}$ simultaneously. They are highly disfavored even with present data, Fig.[1-7]. The reason lies in the fact that string models with early-time effects and inflationary cosmology with modified primordial spectrum offer only one source for seeding all $TT,TE,EE$ spectra and LSS clustering. It follows that modifying the primordial source by adding features to the inflaton potential, will simultaneously produce the same effect on all other spectra $C^{TT}, C^{TE}, C^{TL}$. A suppression on $C^{TT}$ will also impose a suppression of $C^{TE}, C^{LL}$. Clearly this prediction of conventional single-field inflationary physics is already disfavored by WMAP data, Fig.[1-5].While the $EE-$spectrum is useful in discriminating models, it can not provide conclusive evidence due to degeneracies among cosmic priors as well as the limitations imposed by cosmic variance. But upcoming weak lensing surveys will provide strong evidence through data for $C^{TL}$ in the existence of more than the primordial spectrum degrees of freedom. If there is a cuttoff on the inflaton primordial spectrum then correlation $C^{TL}$ should be of order one and $C^{EE}$ spectrum highly suppressed at the cuttoff scale. String theory can provide a richer 'pool'of sources for the origin of perturbations and clustering at large scales. It is this fact that contains the unique handle for our method of detection of imprints for string theory.
Multi-field inflationary models were not included in this analysis. After this work appeared \cite{steenlaura}, authors of\cite{waynehu} extended this analysis to multifield inflation. By aplying various known constraints on isocurvature and multifield inflation scale, they reach similar results to ours about the (dis)agreement of multifield inflationary models with presently available data at large scales.

On the other hand, it is interesting that none of the popular stringy models analyzed, Fig[1-7], are a perfect match to data. Nevertheless this does not affect  any of the conclusions in this work. The purpose of this analysis was not to advocate or rule out specific models but rather to offer a new method which can make a generic model-independent prediction whether evidence for new physics exists. Once, upcoming data from weak lensing surveys and WMAP becomes available then this method can be also used to analyse models on a case by case basis.
 
With the current data available for the acceleration of the universe and the combined data of CMB+LSS, our result is that the best fit are models that contain two modified sources: a late-time modified Friedman equation which uniquely alters the ISW effect; as well as a  modified perturbation source which changes the clustering properties at large scales.A generic prediction for this type of string modifications is the lack of correlations between $T,L$. If weak lensing data finds a suppressed $C^{TL}$ spectrum at large scales than this is strong evidence for the existence of two different sources for $T$ and $L$. Models with modified gravity on very large scales and modified clustering look promising. After all it may not be a coincidence that the accceleration of the universe and CMB power suppression occur at the same scale. 

Future large scale cosmic shear measurements should be able to see a clear difference between the stringy models and the conventional $\Lambda$CDM models, with or without modified spectra. Therefore weak lensing at large scales has the potential to provide new important information about our universe. An improvement of $C^{TE}$ data expected in a near future from WMAP will also allow us to better discriminate among various models and mechanisms. With data available in the near future, such as polarization measurements of the CMB by WMAP, and weak lensing surveys, it will be possible to exclude many of the models and collect further evidence for new physics. Using the proposal described here will soon provide a more definite answer to the question: {\it Do we have observational evidence for string theory in the sky?}.


\begin{thebibliography}{99}

\bibitem{de}  P.M. Garnavich et al, Ap.J. Letters \textbf{493}, L53-57
(1998); S. Perlmutter et al, Ap. J. \textbf{483}, 565 (1997); S.
Perlmutter et al (The Supernova Cosmology Project), Nature \textbf{391} 51
(1998); A.G. Riess et al, Ap. J. \textbf{116}, 1009 (1998);
A.G.Riess {\it et al.}  [Supernova Search Team Collaboration],
Astron.\ J.\  {\bf 116}, 1009 (1998) [arXiv:astro-ph/9805201];
S.Perlmutter {\it et al.}  [Supernova Cosmology Project Collaboration],
Astrophys.\ J.\  {\bf 517}, 565 (1999) [arXiv:astro-ph/9812133].

\bibitem{wmap} 
C.L.Bennett {\it et al.}, ``First Year Wilkinson Microwave
Anisotropy Probe (WMAP) Observations: Preliminary Maps and Basic
Results,'' Astrophys.\ J.\ Suppl.\  {\bf 148}, 1 (2003),[arXiv:astro-ph/0302207]; Hinshaw, et al.,[arXiv:astro-ph/0302217], Astrophys.J.Suppl. 148 (2003); Tegmark et al, [arXiv:astro-ph/0302496].

\bibitem{steenlaura} S.~Hannestad and L.~Mersini-Houghton, Phys.Rev.D {\bf 71},123504,(2005), [arXiv:hep-ph/0405218].

\bibitem{cosmicvar} C. Skordis, J. Silk  [arXiv:astro-ph/0402474], (2004);M. Kamionkowski and A.Loeb,Phys.Rev.D {\bf 56}, 4511 (1997) [arXiv:astro-ph/9703118];N.Seto and M.Sasaki,Phys.Rev.D {\bf 62}, 123004 (2000) [arXiv:astro-ph/0009222];A.R.Cooray and D.Baumann, Phys.Rev.D {\bf 67}, 063505 (2003) [arXiv:astro-ph/0211095];J.Portsmouth, [arXiv:astro-ph/0402173].
\bibitem{waynehu} W.~Hu and R.~Scranton, Phys.Rev.D.{\bf 70} 123002 (2004),[arXiv: astro-ph/0408456].
\bibitem{contaldi} J.~R.~Bond, C.~R.~Contaldi, A.~M.~Lewis and D.~Pogosyan, Int.J.Theor.Phys. {\bf 43}, 599 (2004), [arXiv: astro-ph/0406195].

\bibitem{cross} O.Dore,G.P.Holder,A.Loeb, [arXiv:astro-ph/0309281], (2003).

\bibitem{kesden} M. Kesden, M. Kamionkowski, A. Cooray, [arXiv:astro-ph/0306597], (2003).

\bibitem{ale} Alessandro Melchiorri and Laura Mersini-Houghton, Matters of Gravity Spring Issue, [arXiv:hep-ph/0403222], (2004).

\bibitem{linde} Contaldi, C.R., Peloso, M., Kofman, L., Linde, A., JCAP 0307:002,(2003), [arXiv:astro-ph/0303636].

\bibitem{easther} Jennifer Adams, Bevan Cresswell,Richard Easther, [arXiv:astro-ph/0102236], (2001)


\bibitem{costa} A.de Oliveira-Costa, M.Tegmark, M.Zaldarriaga and A.Hamilton,
[arXiv:astro-ph/0307282].
\bibitem{efstathiou} G.Efstathiou,
%
[arXiv:astro-ph/0310207];J-P Uzan,U.Kichner,G.F.R.Ellis,[arXiv:astro-ph/0302597].
\bibitem{katie} Bastero-Gil,M.,Freese,K.,Mersini-Houghton,L., Phys.Rev.D68, [arXiv:hep-ph/0306289], (2003).


\bibitem{tp} R.Easther,B.R.Greene,W.H.Kinney and G.Shiu, Phys.Rev.D {bf 64},[arXiv:hep-th/0104102], Phys.Rev.D {\bf 67},[arXiv:hep-th/0110226],Phys.Rev.D {\bf 66},[arXiv''hep-th/0204099];M.Bastero-Gil, P.H.Frampton, L.Mersini, Oct.2001. Phys.Rev.{\bf D65}:106002, [arXiv: hep-th/0110167];
F.~Lizzi, G.~Mangano, G.~Miele and M.~Peloso,
JHEP {\bf 0206}, 049 (2002);
O.Elgaroy and S.Hannestad, Phys.Rev.D68,[arXiv:astro-ph/0307011];M.Bastero-Gil and L.Mersini,Phys.Rev.D {\bf 65},[arXiv:astro-ph/0107256];A.Kempf and J.C.Niemeyer,[arXiv:astro-ph/0103225];R.Casadio,L.Mersini,Int.J.Mod.Phys.A {bf 19},[arXiv:hep-th/0205271]; A.A.Starobinsky (Landau Institute), I.I.Tkachev (CERN), Jul 2002. JETP Lett.76:235-239, [arXiv:astro-ph/0207572]; R.Maartens (Portsmouth U., ICG), Living Rev.Rel.7:1-99,2004, [arXiv:gr-qc/0312059]
\bibitem{bgreene} B.~Greene, K.~Schalm., J.~P.~van der Schaar and G.~Shiu, [arXiv: astro-ph/0503458].

\bibitem{ma} Ma and Bertschinger, [arXiv:astro-ph/9506072], (1995).

\bibitem{rs} L.Randall and R.Sundrum, Phys.Rev.Lett. {\bf 83} (1999) 4690; L.Randall and R.Sundrum, Phys.Rev.Lett. {\bf 83} (1999) 3370.
\bibitem{stoicatye} G.Shiu and S.H.Tye, Phys.Lett {\bf B516}. (2001); G.R.Dvali and S.H.Tye Phys.Lett. {\bf B450} (1999).

\bibitem{selftune} 
N.Arkani-Hamed,S.Dimopoulos,N.Kaloper and R.Sundrum,Phys.Lett.B {\bf 480},[arXiv:hep-th/0001197];S.Kachru,M.Schulz and E.Silverstein,Phys.Rev,D62,[arXiv:hep-th/0001206],[arXiv:hepth/0002121];S.M.Carroll and L.Mersini, Phys.Rev.D64,[arXiv:hep-th/0105007].

\bibitem{landscape} L.~Mersini-Houghton, Class.Quant.Grav.{\bf 22}, 3481 (2005), [arXiv:hep-th/0504026]; A.~Kobakhidze and L.~Mersini-Houghton, [arXiv: hep-th/0410213]. 

\bibitem{kogan}  Ian I.Kogan,Stavros Mouslopoulos,Antonion Papazoglou, Phys.Lett.{bf B503}, [arXiv:hep-th/0011138], (2001).

\bibitem{gregory} Ruth Gregory,Valery A.Rubakov and Sergei Sibiryakov, Phys.Rev.Lett.{\bf 84},[arXiv:hep-th/0002072], (2000). 

\bibitem{Deffayet:2001pu}
C.Deffayet, G.R.Dvali and G.Gabadadze,
Phys.\ Rev.\ D {\bf 65}, 044023 (2002)
[arXiv:astro-ph/0105068].

\bibitem{Deffayet:2002sp}
C.~Deffayet, S.~J.~Landau, J.~Raux, M.~Zaldarriaga and P.~Astier,
Phys.\ Rev.\ D {\bf 66}, 024019 (2002)
[arXiv:astro-ph/0201164].


\bibitem{5force} Sean M.Carroll, Phys.Rev.Lett. {bf 81}, [arXiv:astro-ph/9806099], (1998)

\bibitem{balaji}
K.~R.~S.~Balaji, R.~H.~Brandenberger and D.~A.~Easson,
JCAP {\bf 0312}, 008 (2003)
[arXiv:hep-ph/0310368].



\bibitem{dvalikofman} Lev Kofman, [arXiv:astro-ph/0303614], (2003); Gia Dvali, Andrei Gruzinov and Matias Zaldarriaga, Phys.Rev.D 69, [arXiv:astro-ph/0303591], (2003).





\bibitem{spergel} Knop, R.A.et al.(2003), [arXiv: astro-ph/0309368]; Spergel, D.N. et al.(2003), ApJS148, 175; Verde, L. et al., MNRAS335, 432.

\bibitem{melchiorri} Melchiorri,A.,Mersini,L.,Odman,C.,Trodden,M. (2003),Phys.Rev.D68,43509,[arXiv:astro-ph/0211522]; S.Hannestad and E.Mortsell,
Phys.\ Rev.\ D {\bf 66} (2002) 063508
[arXiv:astro-ph/0205096].


\bibitem{desean} Sean M.Carroll, [arXiv:astro-ph/0310342], (2003).


\bibitem{netterfield}
C.~B.~Netterfield {\it et al.}  [Boomerang Collaboration],
Astrophys.\ J.\  {\bf 571} (2002) 604
[arXiv:astro-ph/0104460].

\bibitem{cardassian} Katherine Freese and Matthew Lewis Phy.Lett.{\bf B540}, [arXiv:astro-ph/0201229], (2002); Katherine Freese,   Nucl.Phyd.Proc.Suppl.124, [arXiv:hep-ph/0208264], (2002).

\bibitem{gondolo} P.Gondolo and K.Freese, Phys.Rev.D68, [arXiv:hep-ph/0209322], (2002); P.Gondolo and K.Freese, [arXiv:hep-ph/0211397], (2002); Y.Wang, K.Freese, P.Gondolo, M.Lewis, Astrophys.J. 594, [arXiv:astro-ph/0302064], (2003)

\bibitem{dvalistarkman} A.Lue,G.Starkman,Phys.Rev.D67,[arXiv:astro-ph/0212083];A.Lue,R.Scoccimarro,G.Starkman,[arXiv:astro-ph/0401515].




\bibitem{dvaliturner}
G.~Dvali and M.~S.~Turner,
[arXiv:astro-ph/0301510].


\bibitem{dgp}  G.R.Dvali, G.Gabadadze, M.Porrati, Phys.Lett.B {\bf 485}, [arXiv:hep-th/0005016].






\bibitem{dvali} Dvali, G.,Kachru, S., [arXiv:hep-th/0309095], (2003).


\bibitem{marteens} Tsujikawa,S., Singh, P., Maartens, R.,
[arXiv:astro-ph/0311015], (2003).


\bibitem{nolta} M.~R.~Nolta {\it et al.},
[arXiv:astro-ph/0305097].





\bibitem{Tegmark:2003ud}
M.~Tegmark {\it et al.}  [SDSS Collaboration], ``Cosmological
parameters from SDSS and WMAP,'' [arXiv:astro-ph/0310723].



\bibitem{Verde:2003ey}
L.~Verde {\it et al.}, ``First Year Wilkinson Microwave Anisotropy
Probe (WMAP) Observations: Parameter Estimation Methodology,''
Astrophys.\ J.\ Suppl.\  {\bf 148}, 195 (2003) [arXiv:astro-ph/0302218].

\bibitem{Spergel:2003cb}
D.~N.~Spergel {\it et al.}, ``First Year Wilkinson Microwave
Anisotropy Probe (WMAP) Observations: Determination of
Cosmological Parameters,'' Astrophys.\ J.\ Suppl.\  {\bf 148}, 175
(2003) [arXiv:astro-ph/0302209].

\bibitem{Peiris:2003ff}
H.~V.~Peiris {\it et al.}, ``First year Wilkinson Microwave
Anisotropy Probe (WMAP) observations:  Implications for
inflation,'' Astrophys.\ J.\ Suppl.\  {\bf 148}, 213 (2003)
[arXiv:astro-ph/0302225].

\bibitem{Leach:2003us}
S.M.Leach and A.R.Liddle, 
[arXiv:astro-ph/0306305].

\bibitem{Tegmark:2003uf}
M.~Tegmark {\it et al.}  [SDSS Collaboration],
``The 3D power spectrum of galaxies from the SDSS,''
[arXiv:astro-ph/0310725].

\bibitem{2dFGRS}
M.Colless {\it et al.}, ``The 2dF Galaxy Redshift Survey: Final
Data Release,'' [arXiv:astro-ph/0306581].

\bibitem{Bridle:2003sa}
S.L.Bridle,A.M.Lewis,J.Weller and G.Efstathiou,
Mon.\ Not.\ Roy.\ Astron.\ Soc.\  {\bf 342}, L72 (2003) [arXiv:astro-ph/0302306].


\bibitem{Wang:2000js}
Y.~Wang and G.~Mathews,
Astrophys.\ J.\  {\bf 573}, 1 (2002) [arXiv:astro-ph/0011351].


\bibitem{Kogo:2003yb}
N.~Kogo, M.~Matsumiya, M.~Sasaki and J.~Yokoyama, 
 [arXiv:astro-ph/0309662].

\bibitem{Matsumiya:2002tx}
M.~Matsumiya, M.~Sasaki and J.~Yokoyama,
JCAP {\bf 0302}, 003 (2003) [arXiv:astro-ph/0210365].


\bibitem{Mukherjee:2003cz}
P.~Mukherjee and Y.~Wang, 
Astrophys.\ J.\  {\bf 593}, 38(2003) [arXiv:astro-ph/0301058].

\bibitem{Mukherjee:2003yx}
P.~Mukherjee and Y.~Wang, 
Astrophys.\ J.\  {\bf 598}, 779 (2003) [arXiv:astro-ph/0301562].

\bibitem{Mukherjee:2003ag}
P.~Mukherjee and Y.~Wang, 
Astrophys.\ J.\  {\bf 599}, 1 (2003) [arXiv:astro-ph/0303211].


\bibitem{Cline:2003ve}
J.~M.~Cline, P.~Crotty and J.~Lesgourgues,
JCAP {\bf 0309}, 010 (2003) [arXiv:astro-ph/0304558].


\bibitem{Feng:2003zu}
B.~Feng and X.~Zhang, 
Phys.\ Lett.\ B {\bf 570}, 145 (2003) [arXiv:astro-ph/0305020].


\bibitem{Efstathiou:2003wr}
G.~Efstathiou, 
[arXiv:astro-ph/0306431].


\bibitem{freedman}W.~L.~Freedman {\it et al.},
Astrophys.\ J.\ Lett.\ {\bf 553}, 47 (2001).

\bibitem{dispersion} J. C. Mather et al., Astrophys. J. {\bf 420}, 439 (1994).

\bibitem{elgaroy}
O.Elgaroy and T.Multamaki,
arXiv:astro-ph/0404402.

\bibitem{starobinskyvarun} U.Alam,V.Sahni,T.D.Saini and A.A.Starobinsky, [arXiv:astro=ph/0302302]

\end{thebibliography}
\end{document}